\def\year{2022}\relax
\documentclass[letterpaper]{article} % DO NOT CHANGE THIS
\usepackage{aaai22}  % DO NOT CHANGE THIS
\usepackage{times}  % DO NOT CHANGE THIS
\usepackage{helvet}  % DO NOT CHANGE THIS
\usepackage{courier}  % DO NOT CHANGE THIS
\usepackage[hyphens]{url}  % DO NOT CHANGE THIS
\usepackage{graphicx} % DO NOT CHANGE THIS
\usepackage{natbib}  % DO NOT CHANGE THIS AND DO NOT ADD ANY OPTIONS TO IT
\usepackage{caption} % DO NOT CHANGE THIS AND DO NOT ADD ANY OPTIONS TO IT
\DeclareCaptionStyle{ruled}{labelfont=normalfont,labelsep=colon,strut=off} % DO NOT CHANGE THIS
\usepackage{cite}
\usepackage{amsmath,amssymb,amsfonts}
\usepackage{textcomp}

\usepackage{algorithm}
\usepackage{algorithmic}

\usepackage{newfloat}
\usepackage{listings}
\floatstyle{ruled}
\newfloat{listing}{tb}{lst}{}
\floatname{listing}{Listing}
\title{GenAD: General Representations of Multivariate Time Series for~Anomaly~ Detection}
\author{
    Xiaolei Hua\textsuperscript{\rm 1,5},
    Lin Zhu\textsuperscript{\rm 1,5}\thanks{Lin Zhu is the corresponding author.},
    Shenglin Zhang\textsuperscript{\rm 2,5},
    Zeyan Li\textsuperscript{\rm 3,5},
    Su Wang\textsuperscript{\rm 4},\\
    Dong Zhou\textsuperscript{\rm 4},
    Shuo Wang\textsuperscript{\rm 4},
    Chao Deng\textsuperscript{\rm 1,5}
}
\affiliations{
    \textsuperscript{\rm 1}JIUTIAN Team, China Mobile Research Institute, Beijing, China.
    \textsuperscript{\rm 2}Nankai University, Tianjin, China.\\
    \textsuperscript{\rm 3}Tsinghua University, Beijing, China.
    \textsuperscript{\rm 4}Beijing University of Posts and Telecommunications, Beijing, China.\\
    \textsuperscript{\rm 5}Tsinghua University-China Mobile Communications Group Co., Ltd. Joint Institute, Beijing, China.\\
    \{huaxiaolei, zhulinyj\}@chinamobile.com,
    zhangsl@nankai.edu.cn,
    zy-li18@mails.tsinghua.edu.cn,\\
    \{wangsu, hardworkzd, shuowangg\}@bupt.edu.cn,
    dengchao@chinamobile.com
}

\usepackage{bibentry}
\begin{document}

\maketitle

\begin{abstract}
The reliability of wireless base stations in China Mobile is of vital importance, because the cell phone users are connected to the stations and the behaviors of the stations are directly related to user experience. Although the monitoring of the station behaviors can be realized by anomaly detection on multivariate time series, due to complex correlations and various temporal patterns of multivariate series in large-scale stations, building a general unsupervised anomaly detection model with a higher F1-score remains a challenging task. In this paper, we propose a General representation of multivariate time series for Anomaly Detection(GenAD). First, we pre-train a general model on large-scale wireless base stations with self-supervision, which can be easily transferred to a specific station anomaly detection with a small amount of training data. Second, we employ Multi-Correlation Attention and Time-Series Attention to represent the correlations and temporal patterns of the stations. With the above innovations, GenAD increases F1-score by total 9\% on real-world datasets in China Mobile, while the performance does not significantly degrade on public datasets with only 10\% of the training data. 
\end{abstract}

\section{Introduction}

\begin{figure}[]
\vspace{0.0cm}
	\centering
	\includegraphics[width=3.5in,height=1.7in]{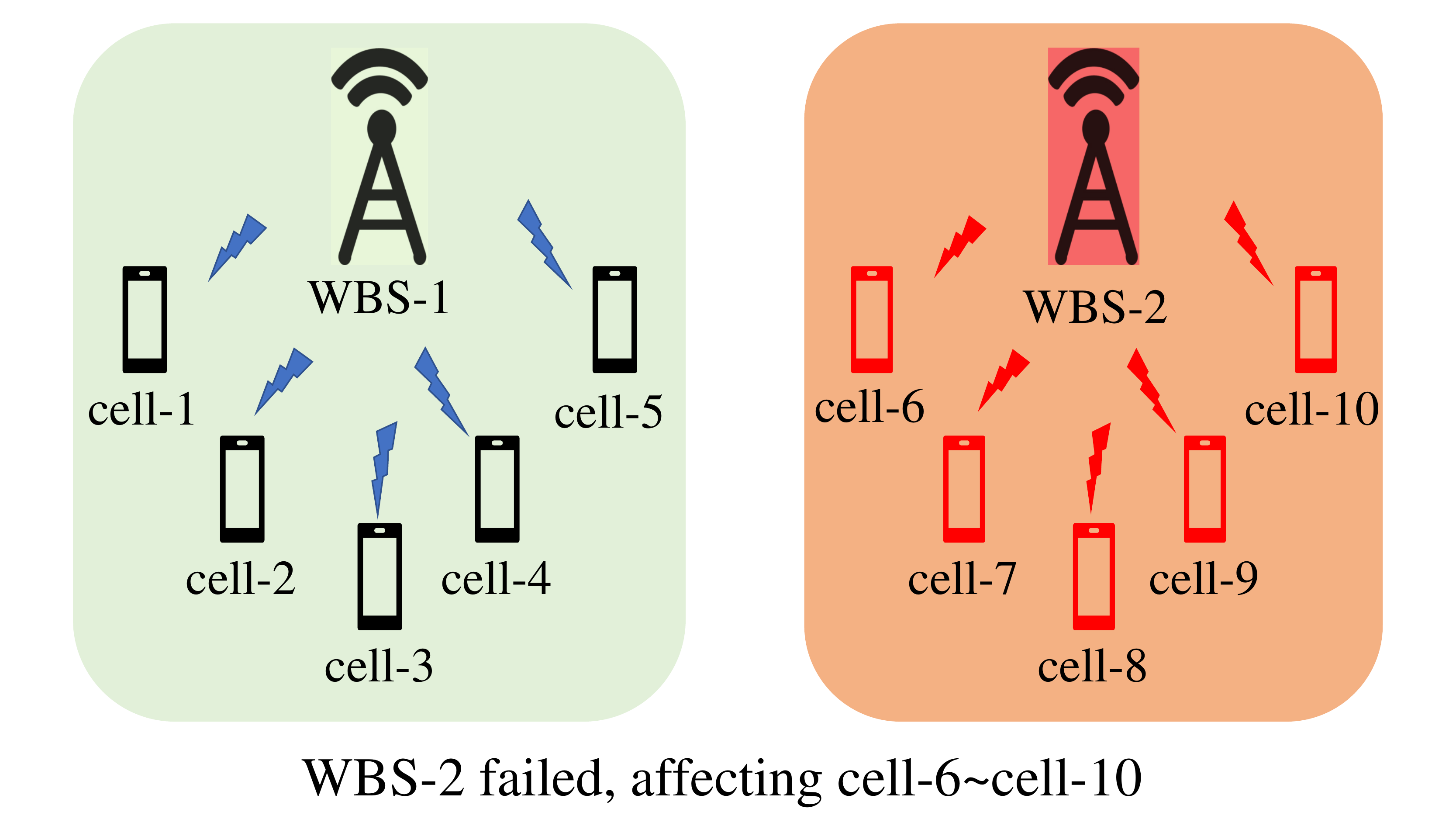}
	\caption{Relationship between WBS and cells.}
	\label{fig:11}
\vspace{0.0cm}
\end{figure}

There are millions of wireless base stations (WBS) in China Mobile, serving hundreds of millions of cell phone users. The reliability of these stations is of vital importance because the users are directly connected to the stations, as shown in Fig. \ref{fig:11}. Once there is an anomaly with WBS-2, it will directly cause various problems to the users connected to it, including failed calls, slow access to the Internet, which will impact user experience and even lead to economic loss.

The behaviors of a WBS can be characterized by the multiple metrics, such as \emph{wireless connection rate}, \emph{wireless drop rate}, \emph{handover success rate}, etc. When one of these metrics becomes anomalous, the WBS is likely to suffer from performance degradation, while in turn impacting user experience. These metrics are continuously collected at a predefined time interval and form a multivariate time series (MTS). Then the detection of WBSes' anomalous behaviors is converted to MTS anomaly detection. Due to anomaly diversity and lack of training labels, only unsupervised methods can be used. Recently, a collection of unsupervised deep learning-based methods\cite{VAE5, VAE6} has been proposed for MTS anomaly detection. But there are two challenges in applying the methods in China Mobile. 
\textbf{Challenge 1: Large-scale WBSes}. 
The number of WBSes in China Mobile has reached a million level, while with the large-scale commercialization of 5G, the number of WBSes is rapidly increasing. Due to the limitations on storage resources for large-scale WBSes, only 480 training data can be used for each WBS. Although deep learning methods \cite{ICLR2018, Chuxu2018A, 2019Robust} perform superiorly for a single device after being carefully designed, it is impossible to deploy such methods on large-scale WBSes in China Mobile due to the huge overhead in training data. Moreover, as millions of stations behave in various patterns because of the various surrounding environment and different manufacturers, it is also inappropriate to train one anomaly detection model for all WBSes, which can not express the heterogeneity of each WBS and degrade the accuracy. Therefore, unsupervised MTS anomaly detection for large-scale and heterogeneous WBSes with a small amount of training data is challenging. 
\textbf{Challenge 2: MTS of each WBS is difficult to express}. The existing methods \cite{ICLR2018, Chuxu2018A, 2019Robust} detect anomalies based on reconstruction errors or reconstruction probabilities, which are directly related to the representation of the correlations and temporal patterns of MTS. But the MTS of each WBS is difficult to express, as there are more than 18 dimensions in the MTS of a WBS, with complex inter-dependencies (i.e., correlations among time series) and various intra-dependencies (i.e., temporal patterns within one time series). More importantly, these inter-dependencies and intra-dependencies of each WBS are dynamic. For example, some WBSes behave a pattern on weekdays, while the pattern can be dynamically changed during the holidays. And an exam or concert can also change the pattern of WBS. These existing methods \cite{VAE1, VAE2, VAE3}, either do not learn the inter-dependencies or intra-dependencies well, or lose the dynamic characteristics.

To address the above challenges, we propose a General multivariate time series representation for Anomaly Detection(GenAD), with the following contributions:

\begin{itemize}
    \item GenAD pre-trains a general model on large-scale WBSes with self-supervision. For each station-specific pattern, only a small amount of data are required to fine-tune the model, which solves the first challenge. 

    \item GenAD employs \emph{Multi-Correlation Attention} and \emph{Time-Series Attention} to represent the complex correlations and temporal patterns of MTS simultaneously. The attention mechanism is introduced to capture the dynamics among time series and within one time series. Multi-head and hidden-layers are also used to capture \emph{Non-linear}, \emph{Coupling}, \emph{Higher-order} correlations, and \emph{Trend}, \emph{Delay}, \emph{Periodicity} of N-dimensional series. 
    
    \item For the large-scale scene, we apply 3000 real-world WBSes in China Mobile for pre-training and the monitoring data of 30 WBSes to evaluate the performance of the general model GenAD (denoted as GenAD(G)). GenAD(G) increases F1-score by total 9\%. To demonstrate the versatility, the performance of GenAD(G) does not significantly degrade on public datasets with only 10\% of the training data.
    
    \item For further demonstrating the representations of complex correlations and various temporal patterns, we train GenAD without pre-training (denoted as GenAD(WP)) for each specific node. GenAD(WP) increases F1-Score by 2.1\% to 4.8\% over the state-of-the-art models on three public datasets, and by 3.4\% on synthetic datasets.

\end{itemize}

\section{BACKGROUND}
\label{BACKGROUND}  
\begin{itemize}
    \item \textbf{WBS}: WBS serves as the central connection point to communicate (making calls, accessing the Internet, etc.) for wireless devices, as shown in Fig. \ref{fig:11}. When a WBS becomes anomalous, users may experience failed calls, slow access to the Internet, resulting in serious degradation of user experience or income loss.
    
    \item \textbf{MTS of WBS}: WBS carries lots of real-time services, so the behaviors of WBS can be characterized by the MTS of the services. For example, \emph{wireless connection rate} can be simply understood as the number of successfully established connections divided by the number of established connection requests for communication, \emph{handover success rate} can be understood as the total number of successful handovers divided by the total number of handover requests. A drop in these metrics indicates that more users suffering from bad experiences.
    
    \item \textbf{General Correlations of MTS in WBS}: The MTS of different WBSes have general correlations. For example, regardless of the environment or the manufacturer of stations, \emph{interference level} directly affects \emph{wireless connection rate}, while there is no direct relationship between \emph{interference level} and \emph{upstream traffic}. The general correlations provide conditions for the general pre-training model.
\end{itemize}

\section{Approach} % General Representations for Anomaly Detection}
\label{headings}
In this section, we first present the problem statement of anomaly detection for MTS.
Second, we introduce the overall workflow and model architecture of GenAD. 
Then, we illustrate \emph{Time-Series Attention} and \emph{Multi-Correlation Attention} in detail, which are the key components of GenAD. 
Finally, we describe the methods for anomaly detection.

\subsection{Problem Statement}

Given the multivariate time series $X$ of a WBS, which contains N dimensions and ${{T}_{X}}$ time points, i.e.,
\begin{equation}
X=({{x}_{1}}^{{{T}_{X}}},{{x}_{2}}^{{{T}_{X}}},\cdot \cdot \cdot ,{{x}_{N}}^{{{T}_{X}}})\in {{\mathbb{R}}^{N\times {{T}_{X}}}}
\label{1}
\end{equation}
we learn the complex correlations and various temporal patterns of $X$, then determine whether there are anomaly segments of $X$ in future time points. 

\subsection{Network Architecture and Pre-training}\label{Network Architecture and Pre-training}  

\begin{figure*}[]
\vspace{0.2cm}
	\centering
	\includegraphics[width=4.3in]{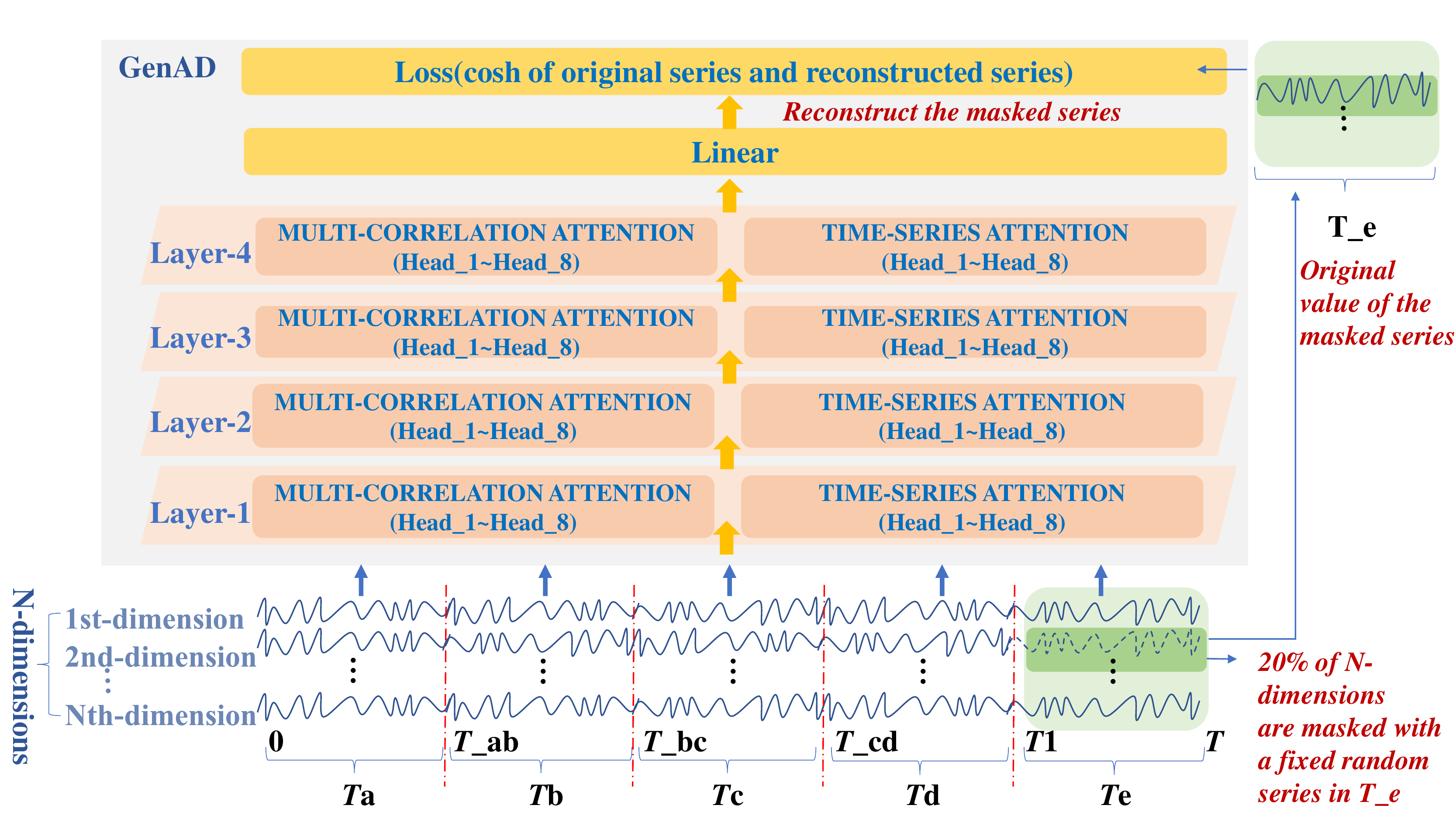}
	\caption{Model Architecture}
	\label{fig:6}
	\vspace{0.2cm}
\end{figure*}

\paragraph{Core ideas}\label{Pre-training methods} 
 GenAD detects anomalies of MTS by evaluating the reconstruction error between the original series and the reconstructed series, so the effect of the reconstructed series determines the effect of anomaly detection. Moreover, in large-scale scenarios of WBS monitoring, there are only 480 data points of each WBS for model training (1/20 of the public SMD Dataset training data size). Then the problem is: how to reconstruct time series well on a small amount of training data? GenAD first randomly selected 20\% of N-dimensions to be masked with a fixed random series in ${T}_{e}$ as shown in Fig. \ref{fig:6} like Bert model\cite{BERT}. Then the left 80\% unmasked series in ${T}_{e}$ and all N-dimensions in ${T}_{a}$ to ${T}_{d}$ are used to reconstruct the 20\% masked series. As 20\% of N-dimensions are selected randomly, GenAD does not know which series of N has been selected to be masked or have been used to be reconstructed during model training, which will force the model to learn correlations and temporal patterns of MTS to minimize loss after the sufficient number of training epochs. Choosing 20\% of MTS in ${T}_{e}$ here means reconstructing one series with four series in each training epoch.

\paragraph{Network Architecture}\label{Network architecture of GenAD} 
Fig. \ref{fig:6} shows network architecture of GenAD. GenAD consists of three parts, which are \emph{four hidden-layers} with \emph{Time-Series Attention} and \emph{Multi-Correlation Attention} in each layer, \emph{Linear-layer} and \emph{Loss function}. \emph{Four hidden-layers} and \emph{Linear-layer} are introduced to reconstruct series. Loss is capture by \emph{cosh} between the reconstructed series and original series. 

\paragraph{More Details of Input Series}\label{More Details of Input Series}    
Input series is N-dimensions MTS with time length ${{T}_{all}}$. Time length ${{T}_{all}}$ is divided into $(0,{{T}_{1}})$ and $({{T}_{1}},T)$, as shown in Fig. \ref{fig:6}, while the length of $(0,{{T}_{1}})$ is equal to 4 times the length of $({{T}_{1}},T)$. Given the length of $({{T}_{1}},T)$ is ${{T}_{e}}$, the length of $(0,{{T}_{1}})$ is sum of ${{T}_{a}}$, ${{T}_{b}}$, ${{T}_{c}}$, and ${{T}_{d}}$, while length of ${{T}_{a}}$, ${{T}_{b}}$, ${{T}_{c}}$, and ${{T}_{d}}$ are the same and all equal to ${{T}_{e}}$. 20\% of N-dimensions are randomly selected to be masked with a fixed random series in ${{T}_{e}}$. Then, the left 80\% original series and 20\% masked series in ${{T}_{e}}$, as well as all series in ${{T}_{a}}$ to ${{T}_{d}}$, are input series of GenAD. 

\paragraph{Pre-training GenAD}\label{Pre-training GenAD}    
We pre-train GenAD based on 3000 real-world WBS Dataset (unlabeled data) with 18-dimensional MTS each station in China Mobile for learning the general correlations and temporal patterns. For each specific station, only a small amount of data are required to fine-tune the model, which empowers GenAD to be used in Large-scale WBSes monitoring.

\subsection{Multi-Correlation Attention and Time-Series Attention}\label{Multi-Correlation Attention and Time-Series Attention} 
\paragraph{Multi-Correlation Attention}\label{Multi-Correlation Attention}   

Existing deep models use CNN, AE, or VAE, to represent correlations of MTS. Once model training has been completed offline, the correlations will not change during online inference. However, the MTS of WBS in China Mobile are not identically distributed. For example, there may exist strong correlation between ${{x}_{i}}$ and ${{x}_{j}}$ of $X$ in $(0,{{T}_{1}})$, but when distribution of ${{x}_{i}}$ and ${{x}_{j}}$ has changed, there will be weak or no correlation between ${{x}_{i}}$ and ${{x}_{j}}$ in $({{T}_{1}},T)$. Therefore, GenAD introduces \emph{Multi-Correlation Attention} to capture the dynamic correlation, as well as non-linear, coupling, and high-order correlations. 

\begin{figure*}[]
\vspace{0.2cm}
	\centering
	\includegraphics[width=4.3in]{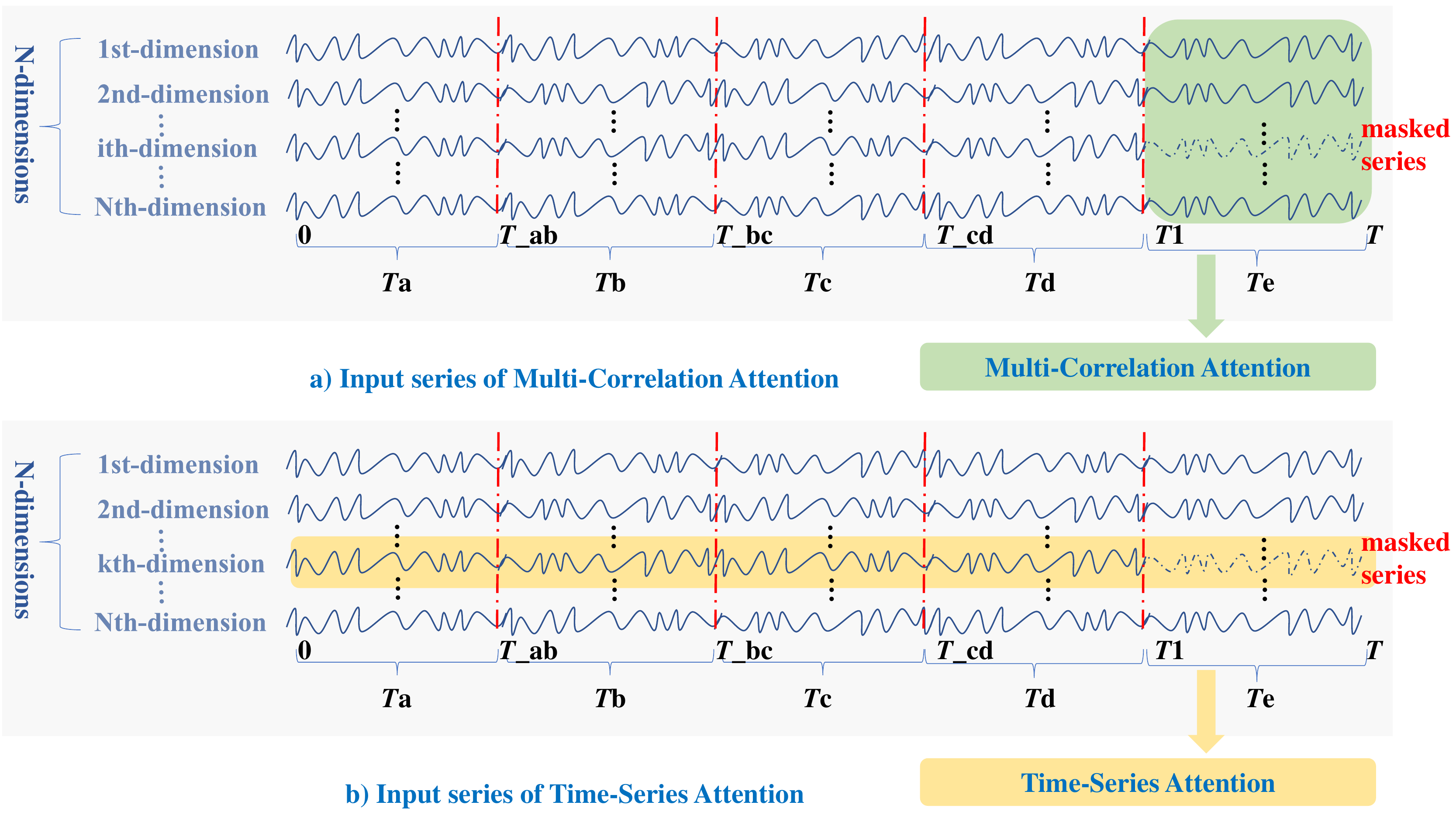}
	\caption{Input of Multi-Correlation Attention and Time-Series Attention}
	\label{fig:10}
	\vspace{0.2cm}
\end{figure*}

The input of \emph{Multi-Correlation Attention} is all the MTS in $({{T}_{1}},T)$ (including the masked and unmasked series), as shown in Fig. \ref{fig:10}a. Assuming the original and masked series in $({{T}_{1}},T)$ is ${{X}^{{{T}_{e}}}}$, and the reconstructed series by surroundings is ${{\tilde{X}}^{{{T}_{e}}}}$ (only the masked series are reconstructed rather than the entire series). Given one of the masked series ${{{x}}_{i}}^{{{T}_{e}}}$ in ${{{X}}^{{{T}_{e}}}}$ as an example, the reconstructed series of ${{{x}}_{i}}^{{{T}_{e}}}$ is ${{\tilde{x}}_{i}}^{{{T}_{e}}}$, 
\emph{Multi-Correlation Attention} is implemented by ``\eqref{12}'':
\begin{small}
\begin{equation}
\begin{aligned}
  & {{{\tilde{x}}}_{i}}^{{{T}_{e}}}\text{=MultiCorrelationAttention}[{{x}_{i}}^{{{T}_{e}}}] \\ 
 & =\sum\limits_{j=(1,2,...,i-1,i+1,...,N)}{soft\max (\frac{({{Q}_{i}}^{{{T}_{e}}}){{({{K}_{j}}^{{{T}_{e}}})}^{Transpose}}}{\sqrt{{{d}}}}){{V}_{j}}^{{{T}_{e}}}} \\ 
\end{aligned}
\label{12}
\end{equation}
\end{small}

where ${{Q}_{i}}^{{{T}_{e}}}$ is obtained through transformation of ${{{x}}_{i}}^{{{T}_{e}}}$ in $({{T}_{1}},{T})$. Transformation of ${{{x}}_{i}}^{{{T}_{e}}}$ means that ${{{x}}_{i}}^{{{T}_{e}}}$ is multiplied by the \emph{transition matrix}, and \emph{transition matrix} can be learned through model training. ${{K}_{j}}^{{{T}_{e}}}$ and ${{V}_{j}}^{{{T}_{e}}}$ are obtained through transforming of surrounding series of ${{{x}}_{i}}^{{{T}_{e}}}$ in $({{T}_{1}},{T})$. \emph{Transpose} is the transpose of a matrix, ${{{{d}}}}$ is the length of ${{{x}}_{i}}^{{{T}_{e}}}$ in time $({{T}_{1}},{T})$. \emph{Multi-Correlation Attention} first obtains the correlations between ${{Q}_{i}}^{{{T}_{e}}}$ and all ${{K}_{j}}^{{{T}_{e}}}$ by $\frac{({{Q}_{i}}^{{{T}_{e}}}){{({{K}_{j}}^{{{T}_{e}}})}^{Transpose}}}{\sqrt{{{d}}}}$, then reconstructes ${{x}_{i}}^{{{T}_{e}}}$ by a weighted sum of ${{V}_{j}}^{{{T}_{e}}}$:
\begin{equation}
{{\tilde{x}}_{i}}^{{{T}_{e}}}\text{=}\sum\limits_{j=(1,2,...,i-1,i+1,...,N)}{si{{m}_{j}}{{V}_{j}}^{{{T}_{e}}}}
\label{13}
\end{equation}
$sim$ is correlations and obtained through \emph{softmax} of the real-time Dot-Product of ${{Q}_{i}}^{{{T}_{e}}}$ and ${{K}_{j}}^{{{T}_{e}}}$. Although \emph{transition matrix} of ${{Q}_{i}}^{{{T}_{e}}}$ and ${{K}_{j}}^{{{T}_{e}}}$ remain constant after offline training, as ${{{x}}_{i}}^{{{T}_{e}}}$ and surrounding series of ${{{x}}_{i}}^{{{T}_{e}}}$ keep changing online, the result of ${{Q}_{i}}^{{{T}_{e}}}$ and ${{K}_{j}}^{{{T}_{e}}}$ also keep changing, with the dynamic correlations ($sim$) between ${{{x}}_{i}}^{{{T}_{e}}}$ and all series being captured, as well as non-linear and coupling correlations from ReLU activation function and Dot-Product operation. The ability to represent correlation can be increased by introducing Multi-head. \emph{Multi-Correlation Attention} also captures the higher-order correlation by stacking multiple layers. In summary, \emph{Multi-Correlation Attention} consists of two layers at least. The first layer learns the dynamic, non-linear, and coupling correlation, and the second layer learns high-order correlations.

\paragraph{Time-Series Attention}\label{Time-Series Attention}
Existing deep models use RNN, GRU, or LSTM, to represent temporal patterns of MTS. However, GRU or LSTM is difficult to capture the various temporal patterns (including dynamic, trend, delay, and periodicity) of N-dimensional series simultaneously, especially when N is large.
GenAD employs \emph{Time-Series Attention} to represent the various temporal patterns of MTS. 

Given ${{{x}}_{k}}$ in ${{{X}}}$ as an example, and assuming ${{{x}}_{k}}$ in time $({{T}_{1}},{T})$ (denoted as ${{{x}}_{k}}^{{{T}_{e}}}$) is masked with a fixed random series. Then the input of \emph{Time-Series Attention} is ${{{x}}_{k}}$ in $(0,T)$, denoted as ${{{x}}_{k}}^{{{T}_{a}}}$, ${{{x}}_{k}}^{{{T}_{b}}}$, ${{{x}}_{k}}^{{{T}_{c}}}$, ${{{x}}_{k}}^{{{T}_{d}}}$ and ${{{x}}_{k}}^{{{T}_{e}}}$, as shown in Fig. \ref{fig:10}b. The reconstructed series of ${{{x}}_{k}}^{{{T}_{e}}}$ is ${{\tilde{x}}_{k}}^{{{T}_{e}}}$ and \emph{Time-Series Attention} is implemented by ``\eqref{18}'' 

\begin{equation}
\begin{aligned}
  & {{{\tilde{x}}}_{k}}^{{{T}_{e}}}\text{=TimeSeriesAttention}[{{x}_{k}}^{{{T}_{e}}}]= \\ 
 & \sum\limits_{t=({{T}_{a}},{{T}_{b}},{{T}_{c}},{{T}_{d}})}{soft\max (\frac{({{Q}_{k}}^{{{T}_{e}}}){{({{K}_{k}}^{t})}^{Transpose}}}{\sqrt{{d}}}){{V}_{k}}^{t}} \\ 
\end{aligned}
\label{18}
\end{equation}

%\[\begin{aligned}
%  & \text{TimeSeriesAttention}[{{x}_{i}}^{{{T}_{e}}}]= \\ 
% & \sum\limits_{t=({{T}_{a}},{{T}_{b}},{{T}_{c}},{{T}_{d}})}{soft\max (\frac{({{Q}_{i}}^{{{T}_{e}}}){{({{K}_{i}}^{t})}^{Transpose}}}{\sqrt{{{d}_{k}}}}){{V}_{i}}^{t}} \\ 
% \label{18}
%\end{aligned}\]

%\[\begin{align}
%  & \text{TimeSeriesAttention}[{{x}_{i}}^{{{T}_{e}}}]= \\ 
% & \sum\limits_{t=({{T}_{a}},{{T}_{b}},{{T}_{c}},{{T}_{d}})}{soft\max (\frac{({{Q}_{i}}^{{{T}_{e}}}){{({{K}_{i}}^{t})}^{Transpose}}}{\sqrt{{{d}_{k}}}}){{V}_{i}}^{t}} \\ 
%\end{align}\]

Where ${{Q}_{k}}^{{{T}_{e}}}$ is the transformation of ${{x}_{k}}$ in time ${{T}_{e}}$; ${{K}_{k}}^{t}$ and ${{V}_{k}}^{t}$ are the transformation of ${{x}_{k}}$ in time ${{T}_{a}}$, ${{T}_{b}}$, ${{T}_{c}}$, ${{T}_{d}}$; \emph{Transpose} is the transpose of a matrix; ${{{{d}}}}$ is the length of ${{x}_{k}}$ in time ${{T}_{e}}$. Equation ``\eqref{18}'' means that GenAD reconstructs ${{x}_{k}}$ in time ${{T}_{e}}$ by learning the various temporal patterns of ${{x}_{k}}$ in ${{T}_{a}}$, ${{T}_{b}}$, ${{T}_{c}}$, ${{T}_{d}}$.

After sufficient model training for Multi-head and Hidden layer of \emph{Time-Series Attention}, GenAD can represent the various temporal patterns of MTS simultaneously. Attention mechanism, different from LSTM, also allows for parallelization, which is especially important at longer series lengths.

\paragraph{Fusion of Two Attentions}\label{Summary of Network Architecture}
GenAD introduces \emph{Time-Series Attention} and \emph{Multi-Correlation Attention} to capture the temporal patterns and correlations of MTS. We can set the first and third \emph{hidden-layers} of GenAD to be \emph{Time-Series Attention}, while the second and fourth \emph{hidden-layers} to be \emph{Multi-Correlation Attention}. However, such an independent architecture will artificially reduce the representation of GenAD, so we adopt the fusion of \emph{Time-Series Attention} and \emph{Multi-Correlation Attention}. Each layer captures both the temporal patterns of series and the correlations among series. The fusion can be implemented by concatenating two kinds of attention or adopting the general attention representations which can support two kinds of attention at the same time. GenAD chooses the latter and automatically learns the representation parameters.

\subsection{Method for Anomaly Detetion} \label{Method for anomaly detetion}
As GenAD evaluates the reconstruction error between the original series and the reconstructed series to detect anomalies, we set a two-level dynamic threshold (denoted as metric-level threshold ${Gate}_{metric}$ and entity-level threshold ${Gate}_{entity})$, which is derived from the anomaly rate. For the metric-level, if reconstruction error of one time series at time $t$ is greater then ${Gate}_{metric}$, the time series is declared as anomalous. For an N-dimensional entity at time $t'$, if there are M time series ($M > {Gate}_{metric}$) that are anomalous, then the entity is declared as anomalous. 

GenAD gets a dynamic anomaly threshold on different datasets by anomaly rate. Assuming that the anomaly rate is $A_R$, the anomaly threshold is $Gate$, and the reconstruction error of single-dimensional or multi-dimensional in ${T}_{e}$ are $E^{{{T}_{e}}}$, then 
\begin{equation}
P(E^{{{T}_{e}}}\le Gate)=1-A_R \label{111}
\end{equation}
Assume that the probability density function of $E^{{{T}_{e}}}$ is $f(e)$, and the probability distribution function $F(e)=\int_{-\infty }^{e}{f(t)dt}$, then ``\eqref{111}'' becomes
\begin{equation}
F(Gate)=\int_{-\infty }^{Gate}{f(t)dt}=1-A_R \label{222}    
\end{equation}
We need to get the probability distribution function $F(e)$ of $E^{{{T}_{e}}}$, and then calculate the dynamic anomaly threshold $Gate$. The simple idea is to obtain $F(e)$ by analyzing the statistical features of $E^{{{T}_{e}}}$, for example, $E^{{{T}_{e}}}$ obeys gaussian-distribution or t-distribution. However, this method is not available in large-scale service or equipment scenarios, as the statistical features of each service are different and there are few data of each service for analysis. Futhermore, OmniAnomaly\cite{2019Robust} applies Extreme Value Theory (EVT)\cite{siffer2017anomaly} to estimate the parameters of the distribution for $E^{{{T}_{e}}}$. However, the complexity of this method is high, which makes it difficult to quickly obtain the anomaly threshold $Gate$. Different from the above methods, we estimate the probability density function $f(e)$ of $E^{{{T}_{e}}}$ based on the data of $E^{{{T}_{e}}}$, 

\begin{equation}
 f(e)\text{=}\sum\limits_{i=1}^{Nu{{m}_{E}}}{e^{{{T}_{e}}}_{gat{{e}_{i}}}}, \\ 
 {e^{{{T}_{e}}}_{gat{{e}_{i}}}}=\left\{ \begin{matrix}
   1,if(e-\Delta e\le {{E}^{{{T}_{e}}}_{i}}< e)  \\
   0,else\begin{matrix}
   \begin{matrix}
   \begin{matrix}
   {} & {}  \\
\end{matrix} & {}  \\
\end{matrix} & {}  \\
\end{matrix}  \\
\end{matrix} \right.
\end{equation}

${{Num}_{E}}$ is the number of $E^{{{T}_{e}}}$, ${{E}^{{{T}_{e}}}_{i}}$ is ${{i}_{th}}$ sample data of $E^{{{T}_{e}}}$, and $\Delta e$ is the sample interval of $E^{{{T}_{e}}}$ for $f(e)$. Then we integrate $f(e)$ to get the probability distribution function $F(e)=\sum\limits_{\min {{E}^{{{T}_{e}}}}}^{e}{f(t)}$, and get the dynamic threshold $Gate$ by ``\eqref{222}''. It is worth noting that in order to reduce the error of parameter estimation through sampled data, we set  $A_R'=\eta+A_R$ and ``\eqref{222}'' becomes $F(Gate)=\int_{-\infty }^{Gate}{f(t)dt}=1-A_R'$ , where $\eta\in[-0.01,0.01]$ is set to maximize the F1-score during the validation period. Subsequent experimental results show that this threshold selection method is simple but effective.

\section{EXPERIMENTS}
In this section, we first introduce the experimental datasets, comparison methods and evaluation metrics. We further show the feasibility of our model in large-scale anomaly detection scenarios. In addition, we also conduct experiments to verify the contributions of each component in GenAD. Finally, we describe the Versatility of GenAD.

\subsection{Experimental Setup}
\paragraph{Datasets}   
We evaluate the performance of GenAD on 3 Datasets.

\textbf{(1) WBS Dataset.}
WBS datasets are collected from real-world WBSes in China Mobile, and are divided into unlabeled data and labeled data. As there are large-scale stations that need to be monitored, only 10-day-long MTS with 15-minute sample intervals (960 data points) were stored for each station. We first randomly select 3000 WBSes (unlabeled data) with 18-dimensional MTS in each station. The 3000 WBSes are lack training labels. Moreover, we randomly select three more areas denoted as WBS-Area1, WBS-Area2, WBS-Area3 respectively, and then randomly select 10 WBSes in each area (labeled data). The MTS of each station is divided into two parts of the same length, of which the training set size is 480 points (too small for existing deep models) and the testing set size is 480 points.

\textbf{(2) Public Datasets.}
We use \textbf{SMD} (Server Machine Dataset) \cite{2019Robust}, \textbf{MSL} (Mars Science Laboratory rover) and \textbf{SMAP} (Soil Moisture Active Passive satellite) for experimental studies. \textbf{(i)} SMD is a 5-week-long dataset collected from a large Internet company, where each observation is equally spaced by 1 minute. SMD is collected from 28 machines classified into three groups (denoted as SMD-1, SMD-2, and SMD-3 respectively). Each machine subset contains approximately 28000 time points, the former 14,000 data points consist of the training set, and the latter forms the testing set. \textbf{(ii)} MSL has 132,046 time points, of which the training set size is 58317 and the testing set size is 73729. Compared with SMD, MSL contains more metrics, a higher anomaly rate and more types of anomalies. \textbf{(iii)} SMAP has 562,798 time points, of which the training set size is 135182 and the testing set size is 427616. 

\textbf{(3) Synthetic Dataset.}  
We also use MSCRED Synthetic Dataset\cite{Chuxu2018A} and our synthetic dataset(GenAD Synthetic Dataset) for empirical studies. \textbf{(i)} MSCRED Synthetic Dataset simulates temporal patterns(Trend, Delay, and Periodicity) by Sin and Cos function. Two sinusoidal waves have high correlation when their frequencies are similar. \textbf{(ii)} GenAD Synthetic Dataset simulates various temporal patterns by Sin, Cos, Sawtooth wave and Square wave, while simulates complex correlations by linear, non-linear and higher-order inter-dependencies.

\paragraph{Baseline Methods}
We compare GenAD with the following baseline methods: LSTM-NDT\cite{hundman2018detecting}, the state-of-the-art unsupervised method OmniAnomaly and MSCRED. Of these baselines, LSTM-NDT applies LSTM for MTS prediction, MSCRED detects anomalies based on reconstruction errors, and OmniAnomaly is based on reconstruction probability.

\paragraph{Evaluation Metrics}
We use 3 metrics of \textbf{Precision}, \textbf{Recall}, and \textbf{ F1-Score} to evaluate the anomaly detection performance of GenAD and baseline methods. In practice, abnormal observations usually appear continuously to form an anomaly segment. Generally, operators care more about whether the anomaly detection model can detect a continuous anomaly segment, rather than finding every anomaly in the segment. Following the suggestion of \cite{2019Robust}, we adopt a point-adjust approach\cite{xu2018unsupervised} to calculate the evaluation metrics, that is, if any observation in the anomaly segment is detected, it is considered that the entire segment is correctly detected.

\paragraph{Model Architecture and Parameters}
All models in the experiment use the same architecture. The hyper-parameter of GenAD includes: MASK ratio (default=20\%), Hidden layers(default=3), Attention heads(default=8) and Length of input series(default=32).

\subsection{Model Description}
We use the general model of GenAD (denoted as GenAD(G)) to show the feasibility in large-scale scenarios. GenAD(G) is first pre-trained on the large-scale dataset, then is fine-tuned for each WBS-specific pattern. We also use GenAD without pre-training (denoted as GenAD (WP)) and GenAD without \emph{Time-Series Attention} (denoted as GenAD (WT)) to evaluate the contributions of components.

\subsection{Overall Performance}

\begin{table*}[htbp]
	\centering
	\caption{Results of GenAD(G) and baselines(WBS Datasets)}
	\renewcommand\arraystretch{1.2} 
	\small %此处写字体大小控制命令
	%\scalebox{0.95}{
	%	\resizebox{145mm}{12mm} 
	%	\specialrule{5mm}{5mm}
	\setlength{\tabcolsep}{1.1mm}{	
		\begin{tabular}{|c|c|c|c|c|c|c|c|c|c|c|c|c|}
			\hline
			\textbf{ } &  \multicolumn{3}{c|}{WBS-Area1}  & \multicolumn{3}{c|}{WBS-Area2} & \multicolumn{3}{c|}{WBS-Area3} & \multicolumn{3}{c|}{Total}\\
			\cline{2-13} 
			Method & Pre   & Rec   & F1    & Pre   & Rec   & F1& Pre   & Rec   & F1 & Pre   & Rec   & F1\\
			\hline
			LSTM-NDT &  0.500    & 0.200     & 0.286    &  1.000 &  0.154    & 0.267 & 0.500    & 0.200     & 0.286 & 0.667 & 0.185 & 0.289\\
			\hline
			MSCRED & 0.599    & 1.000     & 0.749    &  0.603 &  1.000    & 0.753 &  0.608 &  1.000    & 0.757 &0.603 &1.000  & 0.753\\
			\hline
			OmniAnomaly & 0.768    & 1.000     & 0.868    &  0.762 &  1.000    & 0.865 &  0.747 &  1.000    & 0.855 &0.759 &1.000  & 0.863\\
			\hline
			GenAD(G) & \textbf{0.926}  & 1.000  & \textbf{0.962}    &  \textbf{0.932} &  1.000    & \textbf{0.965} &  \textbf{0.872} &  1.000    & \textbf{0.931} & \textbf{0.910} & 1.000  & \textbf{0.953}\\
			\hline
		\end{tabular}%
	}
	
	\label{tabel-23}%
\end{table*}%

It is difficult to deploy existing deep learning models on large-scale WBSes in China Mobile, due to the huge overhead in training data for each WBS-specific pattern. Unlike these models, the general model of GenAD (denoted as GenAD(G)) is pre-trained on 3000 real-world unlabeled WBSes in China Mobile to learn the general representation of MTS. Then for the given labeled WBSes dataset, only 480 data points of each WBS are required to fine-tune the general model. After each WBS is fine-tuned for 10,000 steps, the loss is reduced to below 7, indicating that the model represents the complex correlations and various temporal patterns of WBS Datasets well. Table \ref{tabel-23} lists the precision, recall, and F1-score of different WBSes, in which the best score is highlighted in bold. Note that the precision and recall are the average values of the datasets, and F1 is derived from the precision and recall. Although each of these methods provides an algorithm for calculating the anomaly threshold, they all need a parameter (e.g.  the parameter 'level' in OmniAnomaly) that quantifies the degree of anomaly as input. Therefore, we conduct multiple experiments to choose parameters to get the best results for all the methods. 

We can see that GenAD(G) performs better than all baseline methods, with F1-score increasing by 9\% over the best baseline on the total WBS labeled dataset. Compared with other deep learning models(e.g. MSCRED and OmniAnomaly), GenAD(G) performs better on two steps framework of pre-training and fine-tuning. As 480 data points of each WBS are too small for training deep models, which affects the representations of correlations and temporal patterns. But GenAD(G) is fine-tuned on pre-training model, so the specific models can be easily transferred with a small amount of training data. We also observe that reconstruction-based models perform better than prediction-based models(e.g. LSTM-NDT). This is because the prediction-based method is more sensitive to noise, and some series in WBS are less predictable due to some uncontrollable factors (such as changes in the network environment).

\subsection{Contributions of Components}
To show the effects of three key techniques in GenAD, which are pre-training, \emph{Time-Series Attention} and \emph{Multi-Correlation Attention}, we reconfigure GenAD to create two models: GenAD without pre-training (denoted as GenAD (WP)) and GenAD without \emph{Time-Series Attention} (denoted as GenAD (WT)).

\paragraph{Effect of Pre-training.}   

\begin{figure}[]
\vspace{0.1cm}
	\centering
	\includegraphics[width=3.1in]{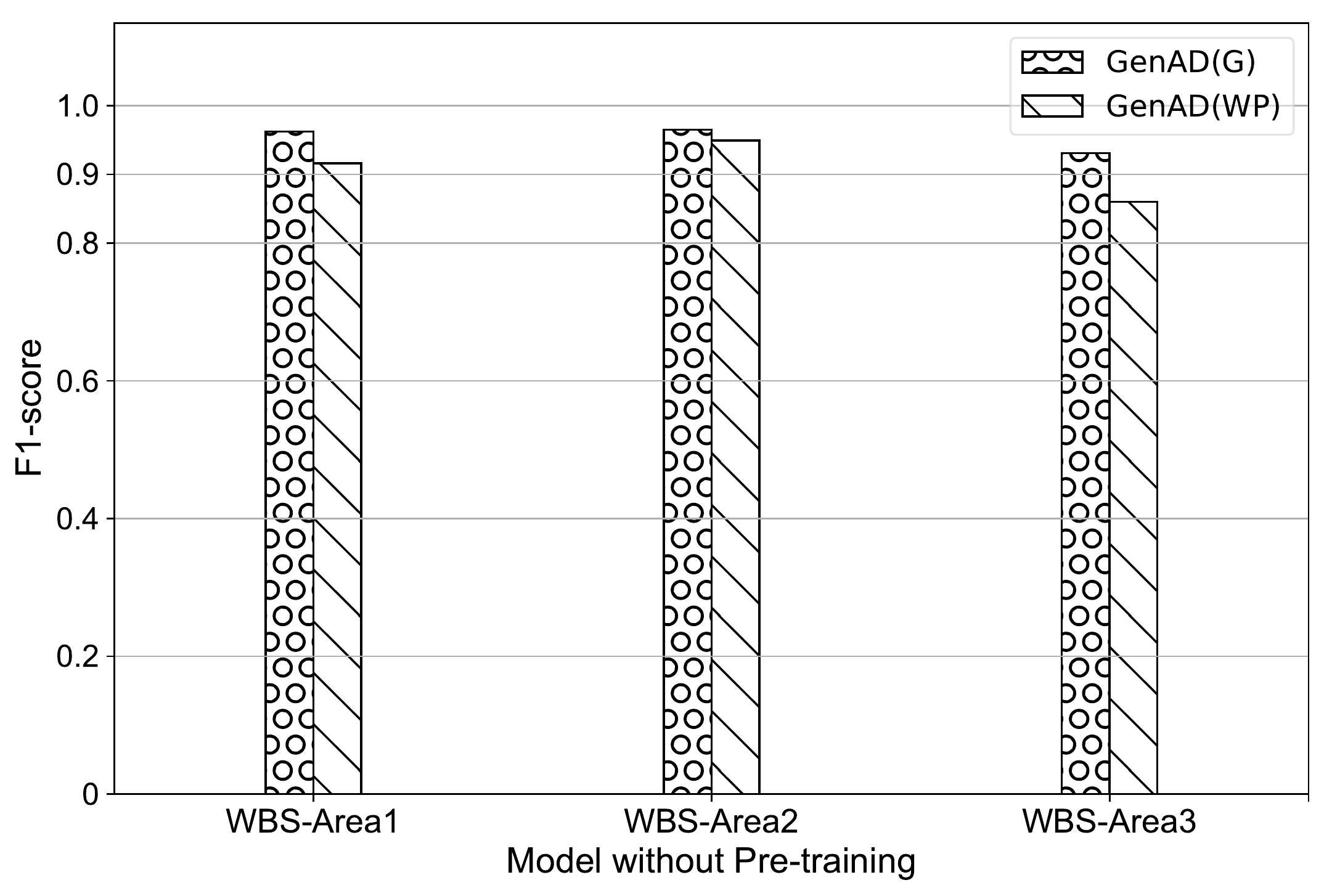}
	\caption{Model without Pre-training}
	\label{fig:12}
	\vspace{0.1cm}
\end{figure}
To verify the effectiveness of pre-training, we train GenAD without pre-training (GenAD(WP)) on WBS datasets. We first train GenAD(WP) on each WBS for 10,000 steps like GenAD(G), but the reconstruction loss is only reduced to around 1000, indicating that GenAD(WP) is not converged well. So we increase the training steps of GenAD(WP) from 10,000 to 20,000, with training time increasing from 8 minutes to 16 minutes. As shown in Fig. \ref{fig:12}, GenAD(G) achieves a higher F1-score on WBS Datasets with only 1/2 training time. This is understandable because GenAD(G) is pre-trained on 3000 real-world WBSes, which empowers GenAD(G) to represent the general correlations and temporal patterns of MTS. After only 10,000 fine-tuning steps, GenAD(G) can be transferred to a specific WBS anomaly detection. But 480 data points and 10,000 training steps of each station are too small to represent the MTS for GenAD(WP).

\paragraph{Effect of Time-Series Attention.}

\begin{figure}[]
\vspace{0.1cm}
	\centering
	\includegraphics[width=3.1in]{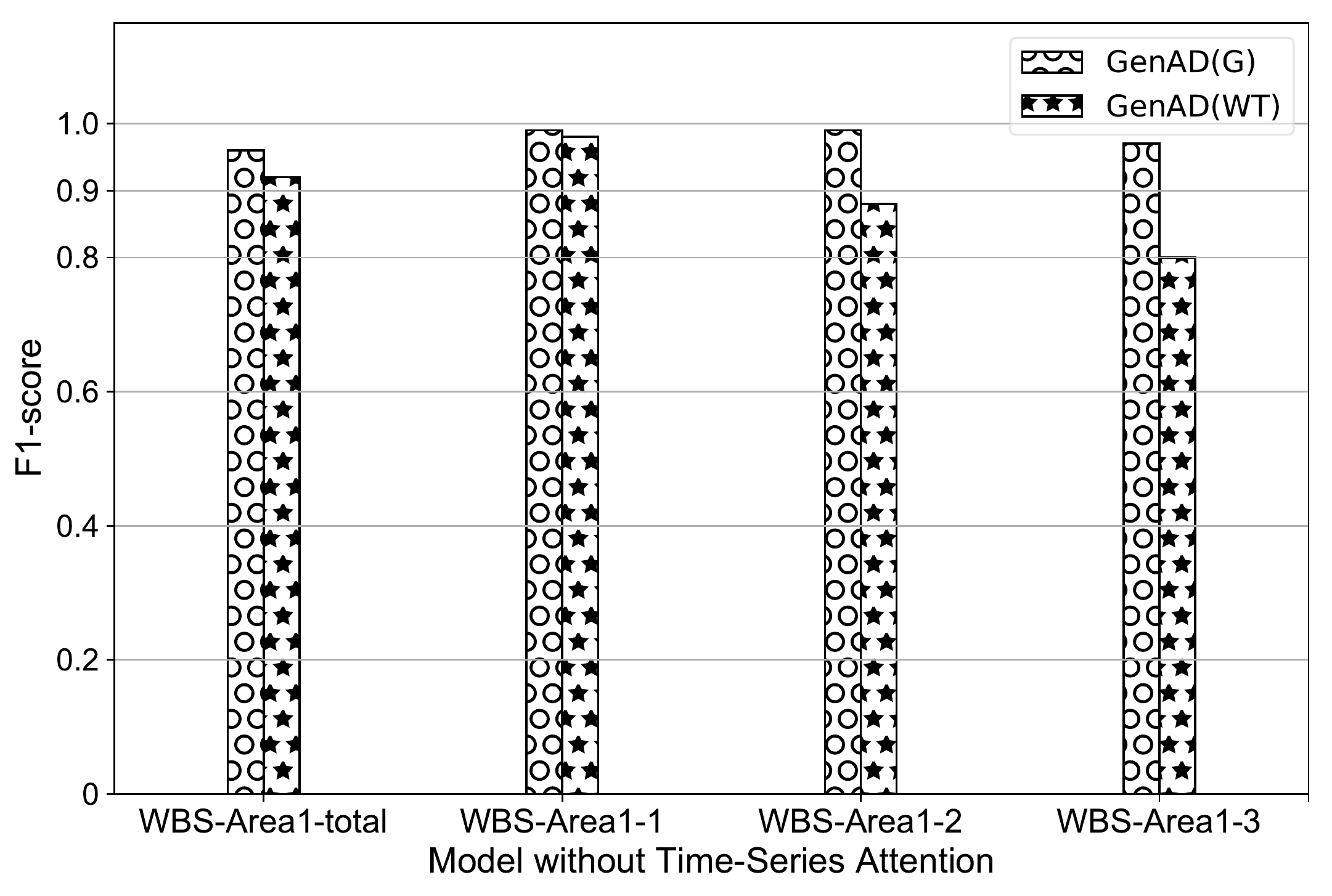}
	\caption{Model without Time-Series Attention}
	\label{fig:7}
	\vspace{0.1cm}
\end{figure}

We evaluate the performance of the GenAD without \emph{Time-Series Attention} (GenAD (WT)) on WBS-Area1, and the results are shown in Fig. \ref{fig:7}, which are WBS-Area1-total and three WBSes selected in WBS-Area1 datasets (denoted as WBS-Area1-1, WBS-Area1-2, and WBS-Area1-3). Due to the loss of various temporal patterns, the total performance (WBS-Area1-total) of GenAD (WT) decreases 4\% compared with GenAD (G). This is because GenAD (WT) can not represent the various temporal patterns of MTS in WBSes well.

\paragraph{Effect of Multi-Correlation Attention.}  
We do not evaluate the performance of GenAD without \emph{Multi-Correlation Attention}. As if we use GenAD without \emph{Multi-Correlation Attention}, the model will detect each time series in isolation and lose the correlations among multivariate series, which results in lower performance of an entity.

\subsection{Versatility of GenAD}

\begin{figure}[]
\vspace{0.1cm}
	\centering
	\includegraphics[width=3.1in]{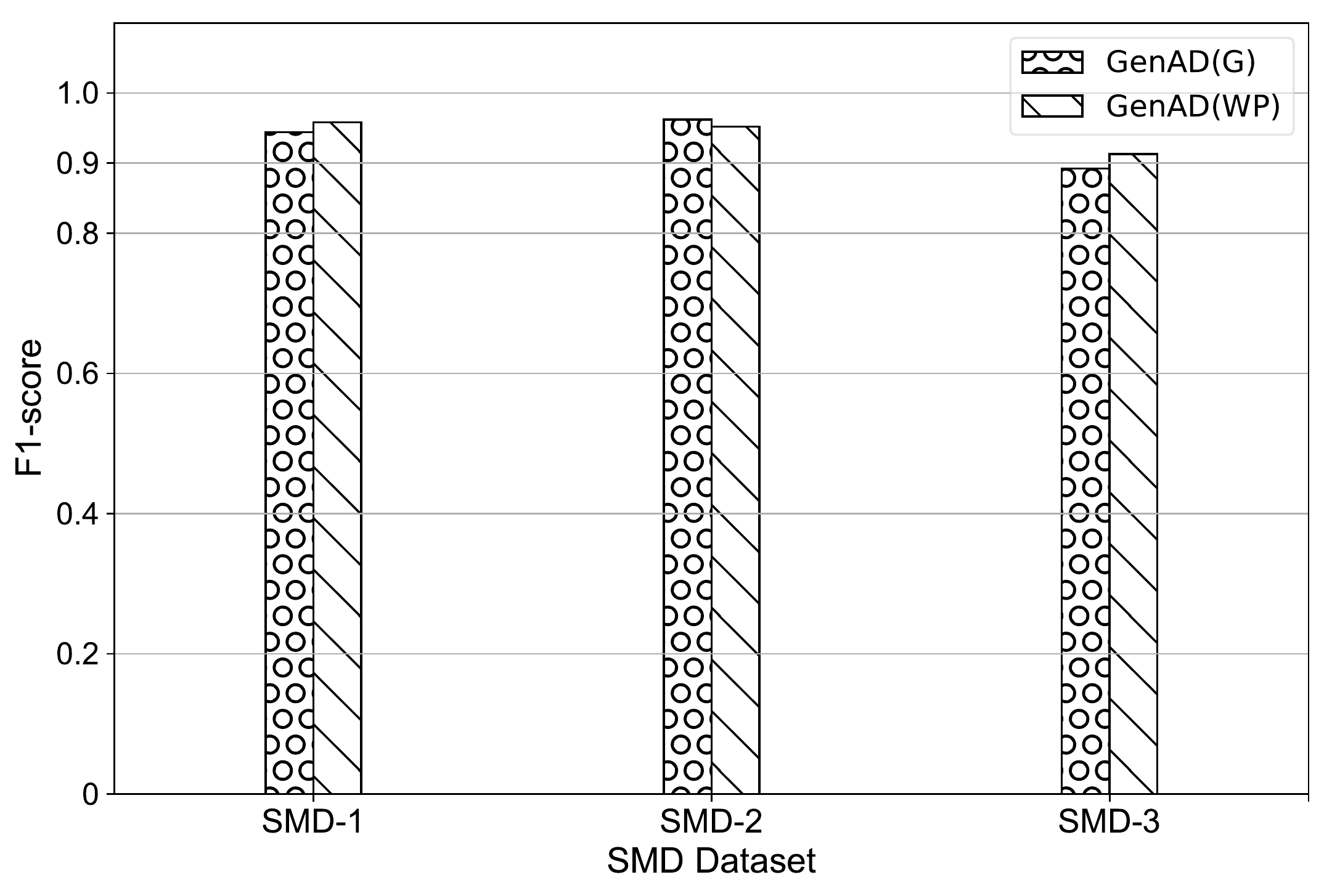}
	\caption{Evaluation of the GenAD(G) on SMD Dataset.}
	\label{fig:1a}
	\vspace{0.1cm}
\end{figure}

\paragraph{General Representation of MTS}   
To demonstrate the versatility of the GenAD(G), we also use SMD Dataset. We first use the data of the first 4 machines in each SMD-1, SMD-2, and SMD-3 (12 machines total) for pre-training, and fine-tune the general model based on only 10\% of the training data in each test machine. Then we train 16 specific models of GenAD (GenAD(WP)) for SMD Dataset. As shown in Fig. \ref{fig:1a}, despite the reduction in training data, the performance of GenAD(G) does not significantly degrade. Interestingly, GenAD(G) performs better than GenAD on SMD-2. This is understandable because GenAD may be over-fitted on some specific training sets and thus sensitive to noise. On the contrary, GenAD(G) obtains fewer details of specific data, which can prevent overfitting and get a higher F1-score. 

\paragraph{Represent Inter-dependencies and Intra-dependencies.}  

%\begin{table}[htbp]
%\caption{Table Type Styles}
%\begin{center}
%\begin{tabular}{|c|c|c|c|}
%\hline
%\textbf{Table}&\multicolumn{3}{|c|}{\textbf{Table Column Head}} \\
%\cline{2-4} 
%\textbf{Head} & \textbf{\textit{Table column subhead}}& \textbf{\textit{Subhead}}& \textbf{\textit{Subhead}} \\
%\hline
%copy& More table copy$^{\mathrm{a}}$& &  \\
%\hline
%\multicolumn{4}{l}{$^{\mathrm{a}}$Sample of a Table footnote.}
%\end{tabular}
%\label{tab1}
%\end{center}
%\end{table}

% Table generated by Excel2LaTeX from sheet '表1'
\begin{table*}[htbp]
	\centering
	\caption{Results of GenAD(WP) and baselines(Public Datasets)}
	\renewcommand\arraystretch{1.2} 
	\small %此处写字体大小控制命令
	%\scalebox{0.95}{
	%	\resizebox{145mm}{12mm} 
	%	\specialrule{5mm}{5mm}
	\setlength{\tabcolsep}{1.1mm}{	
		\begin{tabular}{|c|c|c|c|c|c|c|c|c|c|c|c|c|c|c|c|c|}
			\hline
			\textbf{ } & \multicolumn{3}{c|}{SMD-1} & \multicolumn{3}{c|}{SMD-2} & \multicolumn{3}{c|}{SMD-3} & \multicolumn{3}{c|}{MSL} &  \multicolumn{3}{c|}{SMAP} &  \multicolumn{1}{c|}{Total}\\
			\cline{2-17} 
			Method & Pre   & Rec   & F1    & Pre   & Rec   & F1    & Pre   & Rec   & F1 & Pre   & Rec   & F1 & Pre   & Rec   & F1 & F1\\
			\hline
			LSTM-NDT &  0.773    & 0.298     & 0.431    &  0.767 &  0.282    & 0.412     &  0.819    & 0.428     & 0.557  &   0.632  &    0.330  & 0.433 &  0.525    & 0.776     & 0.627 & 0.528\\
			\hline
			MSCRED & 0.869    & 0.853     & 0.861     & 0.922     & 0.898     & 0.910     & 0.839     & \textbf{0.908}     & 0.872 &  0.892    &   0.934  & 0.912& 0.866    & \textbf{0.982}     & 0.920  & 0.896\\
			\hline
			OmniAnomaly & 0.891    & 0.926    &  0.908    & 0.829    & \textbf{0.994}    & 0.904    & 0.858     & 0.875     & 0.866&   0.916   &   0.854   & 0.884  &0.936  &0.547  &0.690  & 0.862\\
			\hline
			GenAD(WP) & \textbf{0.924}     & \textbf{0.941}     & \textbf{0.933}     & \textbf{0.941}     & 0.964    & \textbf{0.952}     & \textbf{0.884}     & 0.903     & \textbf{0.893} &   \textbf{0.929}   & \textbf{0.966}     &\textbf{0.947}   & \textbf{0.965}  & 0.970  & \textbf{0.968}   & \textbf{0.939}\\
			\hline
		\end{tabular}%
	}
	
	\label{tabel-21}%
\end{table*}%

\begin{table*}[htbp]
	\centering
	\caption{Results of GenAD(WP) and baselines(Synthetic Dataset)}
	\renewcommand\arraystretch{1.2} 
	\small %此处写字体大小控制命令
	%\scalebox{0.95}{
	%	\resizebox{145mm}{12mm} 
	%	\specialrule{5mm}{5mm}
	\setlength{\tabcolsep}{1.1mm}{	
		\begin{tabular}{|c|c|c|c|c|c|c|c|c|c|}
			\hline
			\textbf{ } &  \multicolumn{3}{c|}{MSCRED Synthetic}  & \multicolumn{3}{c|}{GenAD(WP) Synthetic}& \multicolumn{3}{c|}{Total}\\
			\cline{2-10} 
			Method & Pre   & Rec   & F1    & Pre   & Rec   & F1 & Pre   & Rec   & F1\\
			\hline
			LSTM-NDT &  0.200    & 1.000     & 0.333    &  0.750 &  0.600    & 0.667 &0.475 &0.800 &0.596 \\
			\hline
			MSCRED & 0.975    & 0.933     & 0.954    &  0.951 &  0.999    & 0.975  &0.963 &0.966 &0.964 \\
			\hline
			OmniAnomaly & 0.369    & 0.695     & 0.482    &  0.709 &  0.799    & 0.752  & 0.539 & 0.747 & 0.626 \\
			\hline
			GenAD(WP) & \textbf{0.999}  & \textbf{1.000}  & \textbf{0.999}    &  \textbf{0.992} &  \textbf{1.000}    & \textbf{0.996}  & \textbf{0.996} & \textbf{1.000} & \textbf{0.998} \\
			\hline
		\end{tabular}%
	}
	
	\label{tabel-22}%
\end{table*}%

For further demonstrating the representations of complex correlations and various temporal patterns, we train GenAD without pre-training (GenAD(WP)) for each specific node on Public Datasets and Synthetic Dataset.

\textbf{Public Datasets.}
We evaluate GenAD(WP) and baselines on 5 public datasets: SMD-1, SMD-2, SMD-3, MSL and SMAP. Table \ref{tabel-21} shows GenAD(WP) performs better than all baseline methods, with F1-score increasing by 2.1\% to 4.8\% over the best baseline of each dataset and 4.3\% of total datasets. So GenAD(WP) performs better on representations of complex correlations and various temporal patterns.

\textbf{Synthetic Dataset.}
Table \ref{tabel-22} shows GenAD(WP) performs better than all baseline methods, with F1-score increasing by 3.4\% over the best baseline on total Synthetic Dataset. GenAD(WP) represents the various temporal patterns of each time series, including Sin, Cos, Sawtooth wave and Square wave, while represents complex correlations of MTS, including linear, non-linear and higher-order inter-dependencies.

\subsection{Validation of Design Choices}

\paragraph{Impact of number of attention heads and hidden layers}
\begin{table}[htbp]
	\centering
	\caption{F1-score of GenAD(WP) with different number of attention heads and hidden layers}
	\renewcommand\arraystretch{1.2} 
	\small %此处写字体大小控制命令
	%\scalebox{0.95}{
	%	\resizebox{145mm}{12mm} 
	%	\specialrule{5mm}{5mm}

	\begin{tabular}{|c|c|c|c|c|c|}
			\hline
		Attention heads & 12 & 12 & \textbf{12} & 8 & 16 \\
		Hidden layers & 2 & 6 & \textbf{4} & 4 & 4 \\
			\hline
		F1-score & 0.913 & 0.927 & \textbf{0.933} & 0.923 & 0.902 \\
			\hline
	\end{tabular}%

	\label{table 2}%
\end{table}%
The number of attention heads and hidden layers is important for GenAD. Table \ref{table 2} shows the F1-score of GenAD(WP) on SMD-1 by varying different attention head numbers and hidden layer numbers. We observe that by keeping the number of heads constant and changing the number of layers, the 4-layers perform best. The 2-layer \emph{Multi-Correlation Attention} is weak in representing deep and high-order correlations, which leads to a decrease in F1-score; the 6-layer requires a higher amount of training data, and the model convergence is not as good as the 4-layer. In addition, we try the 12-layer attention, and the model can not converge, which further verifies the analysis results. Similarly, keeping 4-layers unchanged, and changing the number of heads, 12-head attention performs best. The 8-head \emph{Multi-Correlation Attention} has a weak ability to capture dynamic, non-linear and coupling relationships, while \emph{Time-Series Attention} has a decline in the ability to capture the number of periodic frequencies and trends, resulting in a decline in F1; 16-heads are consistent with the 6-layer analysis.

\paragraph{How Results Vary If We Change The Ratio of MASK}
If the ratio of MASK selection is increased, for example, modified to 40\%, \emph{Multi-Correlation Attention} will learn less information with less input, which lowers the speed of model training. If the ratio of MASK selection is reduced, for example, modified to 10\%, then each time series can be learned for reconstruction after at least 10 times in model training. While for 20\%, each series can be learned after at least 5 times in model training. So the reduction in the ratio will increase the number of training epochs. Of course, 20\% is not fixed and can be changed according to the actual datasets.

\section{RELATED WORK}
\label{gen_inst}

The existing unsupervised anomaly detection models for MTS can be categorized into the following types:

(1) Anomaly detection can be implemented for each dimension of MTS, and the overall status of an entity is voted or weighed on the outputs of each dimension. Anomaly detection for each dimension can be realized by statistical principles or distance measurements, including 3sigma\cite{3sigma}, boxplot\cite{Boxplot}, HBOS\cite{HBOS}, KNN, AvgKNN\cite{AvgKNN}, OCSVM\cite{das2010multiple}, etc. These models obtain the distribution or the farthest distance of normal series, while anomaly is detected as outliers. These models require less training data, which are suitable for large-scale series. However, most of these models perform better for short-term abnormalities, while the performance will be attenuated for long-term abnormalities. Anomaly detection for each dimension can also be realized by prediction(ARIMA\cite{arima}, LSTM, Prophet\cite{Prophet}) or reconstruction (AE, VAE). These models learn the intra-dependencies of series and adapt well to time series. However, these models detect each time series in isolation and lose the correlations among multivariate series, which results in lower performance of an entity. Moreover, these models require large and different data for training, which limits the application for large-scale series.

(2) Anomaly detection can also be implemented at the Entity-level than for each dimension. PCA\cite{PCA}, RPCA\cite{RPCA}, MCD\cite{MCD} learn Inter-dependency patterns of multivariate series and detect anomalies based on changes in correlations, which requires less training data. But these models only represent the linear correlations. DEC\cite{2015Unsupervised}, DR+K-means\cite{2016Towards}, DAGMM reduce the dimension of multivariate series by DNN or AE, which solves the problem of representations of non-linear correlation. RSRAE\cite{ICLR2020} combines AE and RSR to learn non-linear correlation, which also exhibits robustness to abnormal points in the training data. However, none of the above models is suitable for time series. MSCRED\cite{Chuxu2018A} detected anomalies by calculating the differences between the reconstructed and the original correlation matrix. However, MSCRED only measures the correlation matrix and loses the intra-dependencies of the series itself. Moreover, the correlation matrix is obtained by a simple inner-product of two time series, which is impossible to find the deep higher-order correlation. Omni\cite{2019Robust} introduces VAE to mine the inter-dependencies and combines GRU to represent intra-dependencies of series, which achieves robust results. However, single-layer GRU is difficult to capture the various temporal patterns of N-dimensional series when N is large. Omni also lose dynamic correlation and requires large training data, which also limits the application in large-scale MTS.

Compared with the above approaches, GenAD can not only obtain dynamic, non-linear and deep higher-order correlations, but also represent various temporal patterns of time series. More importantly, GenAD proposes a pre-training algorithm on large-scale MTS, which can be easily transferred to a specific entity with only a small amount of training data.

\section{Conclution}
\label{others}
In this paper, we propose a general pre-training algorithm on large-scale WBSes in China Mobile, which can be easily transferred to a specific AD task with only a small amount of training data. we also adopt \emph{Multi-Correlation Attention} to represent the complex correlations among the MTS and employ \emph{Time-Series Attention} to represent the various temporal patterns of each time series. Through extensive experiments, GenAD increases F1-score by total 9\% on real-world datasets in China Mobile, while the performance does not significantly degrade on public datasets with only 10\% of the training data. We have applied GenAD in monitoring large-scale WBS behaviors in China Mobile, which has improved operation efficiency by 30\%-40\%.

\bibliography{aaai22}

\end{document}